\documentclass[a4paper,fleqn]{cas-dc}

\usepackage[numbers]{natbib}

\def\tsc#1{\csdef{#1}{\textsc{\lowercase{#1}}\xspace}}
\tsc{WGM}
\tsc{QE}
\tsc{EP}
\tsc{PMS}
\tsc{BEC}
\tsc{DE}

\begin{document}
\let\WriteBookmarks\relax
\def\floatpagepagefraction{1}
\def\textpagefraction{.001}
\shorttitle{Superconductivity and Ferromagnetism in (V$_{0.60}$Ti$_{0.40}$)-Gd alloys}
\shortauthors{Sabyasachi Paul  et al.}

\title [mode = title]{Coexisting superconductivity and ferromagnetism in the (V$_{0.60}$Ti$_{0.40}$)-Gd alloys }                      

\author[1,2]{Sabyasachi Paul}
\credit{Investigation, Data curation, Formal analysis}

\address[1]{Free Electron Laser Utilization Laboratory, Raja Ramanna Centre for Advanced Technology, Indore - 452013, India}
\address[2]{Homi Bhabha National Institute, Training School Complex, Anushakti Nagar, Mumbai - 400094, India}

\author[1,2]{SK. Ramjan}
\credit{Investigation,  Data curation}

\author[1]{L. S. Sharath Chandra}[%
orcid=0000-0002-1253-6035]
\cormark[1]
\ead{lsschandra@rrcat.gov.in}
\credit{Conceptualization of this study, Formal analysis, Writing - Original draft preparation}

\author[1,2] {M. K. Chattopadhyay}
 \credit{Methodology, Writing - Review & Editing, Project administration} 

\begin{abstract}
We present here, the effect of microstructure on the magnetic, electrical and thermal properties of (V$_{0.60}$Ti$_{0.40}$)-Gd alloys. The gadolinium is found to be immiscible and precipitates with a size $<$1.2~$\mu$m in the (V$_{0.60}$Ti$_{0.40}$)-Gd alloys. These precipitates enhance the grain boundary density. The (V$_{0.60}$Ti$_{0.40}$)-Gd alloys become ferromagnetic below $T_{mc}$ = 295~K with an increase in the superconducting transition temperature ($T_{sc}$). Though the disorder increases with increasing Gd content, the electronic thermal conductivity ($\kappa_{e} (H = 0)$) reduces by at most 15\% which is in contrast with the 80\% decrease of the phononic thermal conductivity ($\kappa_{l} (H = 0)$). Our analysis suggests that the magnetic moments of Gd precipitates polarize the conduction electrons along and around the grain boundaries leading to increase in the mean free path of the electrons. The partial suppression of spin fluctuations in the (V$_{0.60}$Ti$_{0.40}$)-Gd alloy by the conduction electron polarization enhances the $T_{sc}$.

\end{abstract}

\begin{graphicalabstract}
\includegraphics[width=120 mm, keepaspectratio]{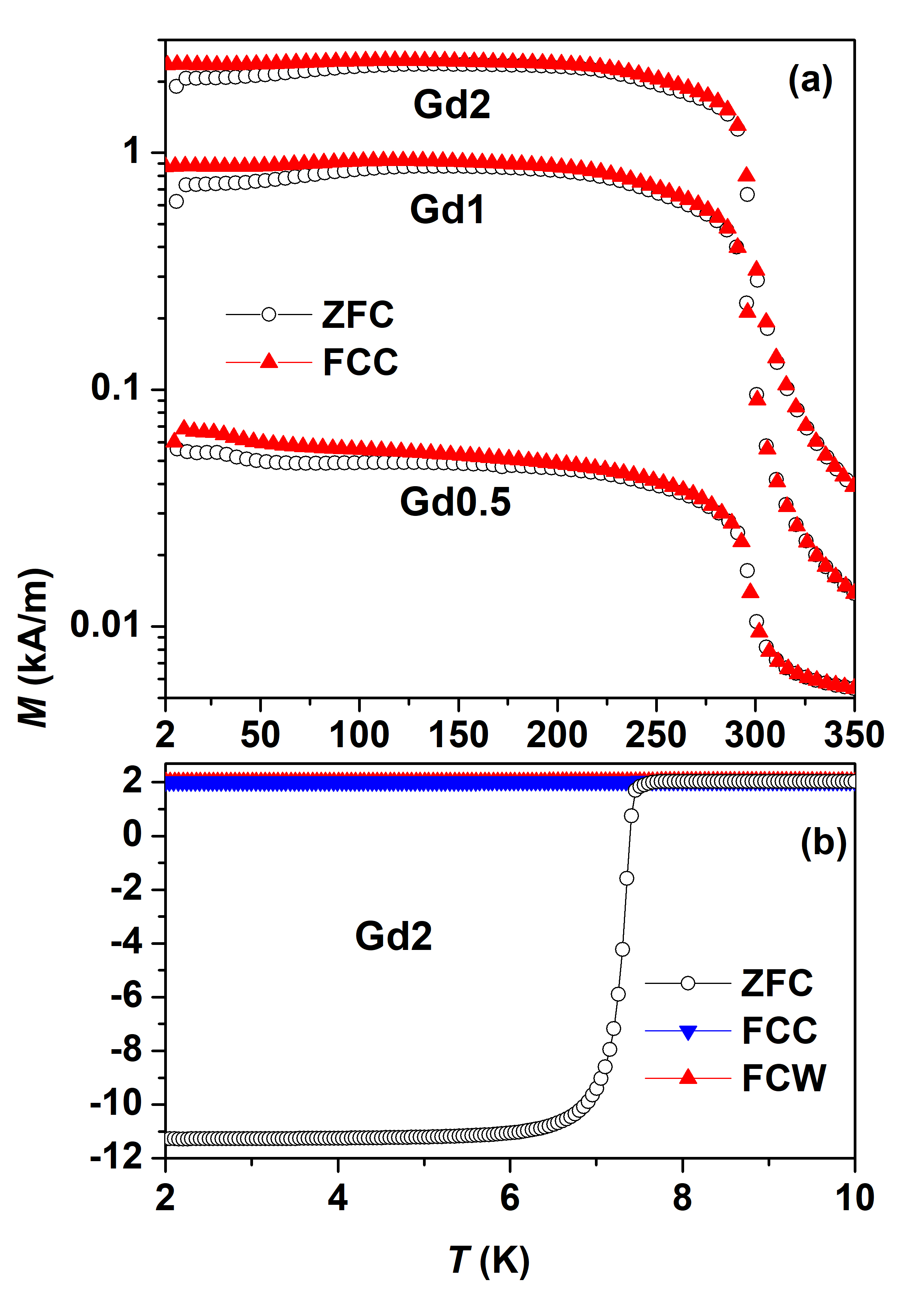}
\end{graphicalabstract}

\begin{highlights}
\item Coexistence of superconductivity and ferromagnetism
\item Conduction electron polarization along and around the grain boundaries
\item Enhancement of electrical and thermal conductivities with increasing disorder
\end{highlights}


\maketitle

\section{Introduction}

The body centred cubic $\beta$-V$_{1-y}$Ti$_y$ alloys are considered to be promising candidates alternate to the Nb-based alloys and compounds for superconducting magnet applications, especially in the neutron radiation environment \cite{mat14, mat13, tai07}. While these alloys are mechanically strong, their critical current density $J_{c}$ (100-500~Amm$^{-2}$ at 4~K and $H$ = 0 [ref.\cite{mat13,mat15}]) is nearly two orders of magnitude smaller than that of the commercial Nb-Ti superconductors (5000-10000~Amm$^{-2}$ at 4~K and $H$ = 0 [ref.\cite{sha19} and the references therein]). Earlier, we have shown that the low $J_c$ of the V$_{1-y}$Ti$_y$ alloys is due to the low grain boundary (GB) density \cite{mat13, mat15}. Rare earth elements are immiscible in the transition metal elements such as titanium and vanadium \cite{cha10, lov60}. Recently, we have shown that the addition of Gd to the V$_{0.60}$Ti$_{0.40}$ alloy increases the GB density \cite{paul20}. These GBs along with the Gd precipitates improve the $J_c$ of the V$_{0.60}$Ti$_{0.40}$ alloy from 200~Amm$^{-2}$ to about 800~Amm$^{-2}$ for 1 at.\% Gd in the V$_{0.60}$Ti$_{0.40}$ alloy \cite{paul20}. The Gd containing  V$_{0.60}$Ti$_{0.40}$ alloys also exhibit:\cite{paul20} {\it\\(i) a ferromagnetic transition at $T_{mc}$ = 295~K, }and \\ {\it (ii) slight enhancement of the superconducting transition temperature} ($T_{sc}$).

The $T_{sc}$ of several binary alloys such as La$_{1-x}$Gd$_x$ \cite{mat58, wol15}, Pb$_{1-x}$Gd$_x$ \cite{wol15, sch60, woo65} etc., and pseudo-binaries such as Ce$_{1-x}$Gd$_x$Ru$_2$ \cite{wol15, mat58a} and La$_{1-x}$Gd$_x$Al$_2$ \cite{wol15, map68} etc., decreases with increasing $x$ (the amount of Gd) because of pair breaking due to exchange interaction between the Gd 4$f$ moments and the conduction electrons \cite{wol15}. The magnetic order in these alloys appear in the samples where the $T_{sc}$ got substantially decreased, or after the complete suppression of superconductivity \cite{mat58, mat58a}. Therefore, it is surprising that the superconducting properties enhance with the onset of ferromagnetism in the Gd containing V$_{0.60}$Ti$_{0.40}$ alloys.  

Earlier, we have shown that the enhancement of $T_{sc}$ in the V$_{y}$Ti$_{1-y}$ alloys in comparison with that of vanadium is due to the reduction in the itinerant ferromagnetic spin fluctuations (IFSF) \cite{mat14, mat14ejpb}. Evidence of ferromagnetic spin fluctuations (FSF) in V$_{y}$Ti$_{1-y}$ manifests as a $T^2$ dependence in the resistivity ($\rho$($T$)) at $T < $100~K along with a negative $T^5$ contribution, and a $T^3$ln$T$ contribution to the heat capacity ($C_p$($T$)) along with an enhanced electronic contribution \cite{mat14, mat14ejpb}. The itinerancy of FSF in these alloys was argued on the basis of a linear field dependence of magnetization ($M(H)$) and the absence of magnetoresistance in the normal state at $T > 3T_{sc}$ \cite{mat14,mat14ejpb}. The $\mu$SR spectroscopy studies by Barker et al., have confirmed the existence of FSF in V$_{0.55}$Ti$_{0.45}$ below 120~K \cite{bar14}. It is reported in literature that the addition of Gd suppresses FSF in the uranium based systems such as UAl$_2$ and UAl$_3$ due to the exchange interaction of localized Gd 4$f$ moments with the conduction electrons \cite{oli93, luc96, bur91}. Therefore, it is important to understand the role of gadolinium in governing the physical properties of the Gd containing V$_{1-y}$Ti$_{y}$ alloys.

 In this article, we show that the Gd precipitates are found along the GBs in the V$_{0.60}$Ti$_{0.40}$ alloys. Experimental results suggest that the Gd moments polarize the conduction electrons along and around the GBs. This polarization results in bulk ferromagnetism below 295~K and in the partial suppression of IFSF  which in turn enhances the $T_{sc}$. 

\section{Experimental details}

The (V$_{0.60}$Ti$_{0.40}$)-Gd alloys were synthesized by melting the constituent elements  (purity $\geq$ 99.8\%) in stoichiometric ratio in a tri-arc furnace under 99.999\% pure Ar atmosphere \cite{paul20}. The nominal compositions are presented in table 1. The samples were flipped and re-melted 5 times to ensure homogeneity. The samples were characterized by X-Ray diffraction (XRD) measurements performed using $\lambda$ = 0.83019~\AA~ radiation from the BL12 beamline of the Indus-2 synchrotron facility \cite{sin00}. The details of metallography is presented in ref.\cite{paul20}. The resistivity, and thermal conductivity of the samples were measured in a 9~T Physical Properties Measurement System (PPMS, Quantum Design, USA). The magnetization measurements were performed in a 7~T Superconducting Quantum Interference Device based Vibrating Sample Magnetometer (MPMS-3 SQUID-VSM, Quantum Design, USA).

\begin{table}[h]
	
	\begin{center}
			
			\caption{Nomenclature and compositions of the present samples}
			\label{tab:table1}
			\begin{tabular}{c|c|c|c} 
				\hline
				\multicolumn{4}{c}{Compositions in at.\%}\\
				\hline
				\hline
				Sample & V&Ti & Gd\\
				
				\hline
				Gd0 & 60 & 40 &0 \\
				Gd0.5 & 59.5& 40&0.5 \\
				
				Gd1 & 59 & 40 &1 \\
				
				Gd2& 58 & 40 &2 \\
				
				\hline
				
			\end{tabular}
			
			
			
	\end{center}
\end{table}

\section{Results and Discussions}
\subsection{Structural properties and phase diagram of (V$_{0.60}$Ti$_{0.40}$)-Gd alloys }

 \begin{figure}[htb]
 	\includegraphics[width=\linewidth, keepaspectratio]{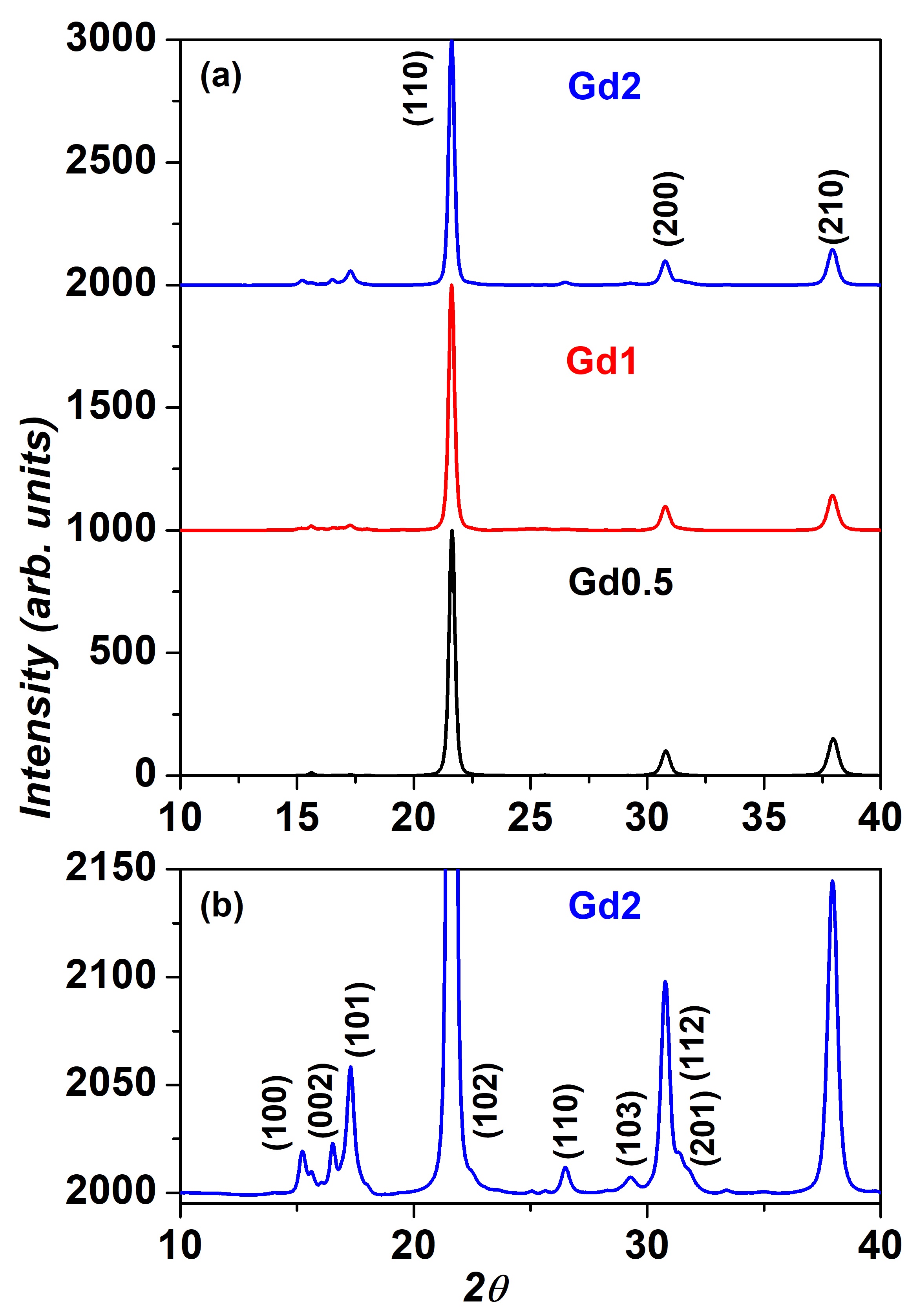}
 	\caption{(a) X-ray diffraction patterns of V$_{0.60}$Ti$_{0.40}$ alloys after the addition of Gd, which is immiscible in the V$_{0.60}$Ti$_{0.40}$ alloys. (b) The reflections corresponding to hcp-Gd.}
	\label{1}
 \end{figure}

 \begin{figure}[htb]
 	\includegraphics[width=\linewidth, keepaspectratio]{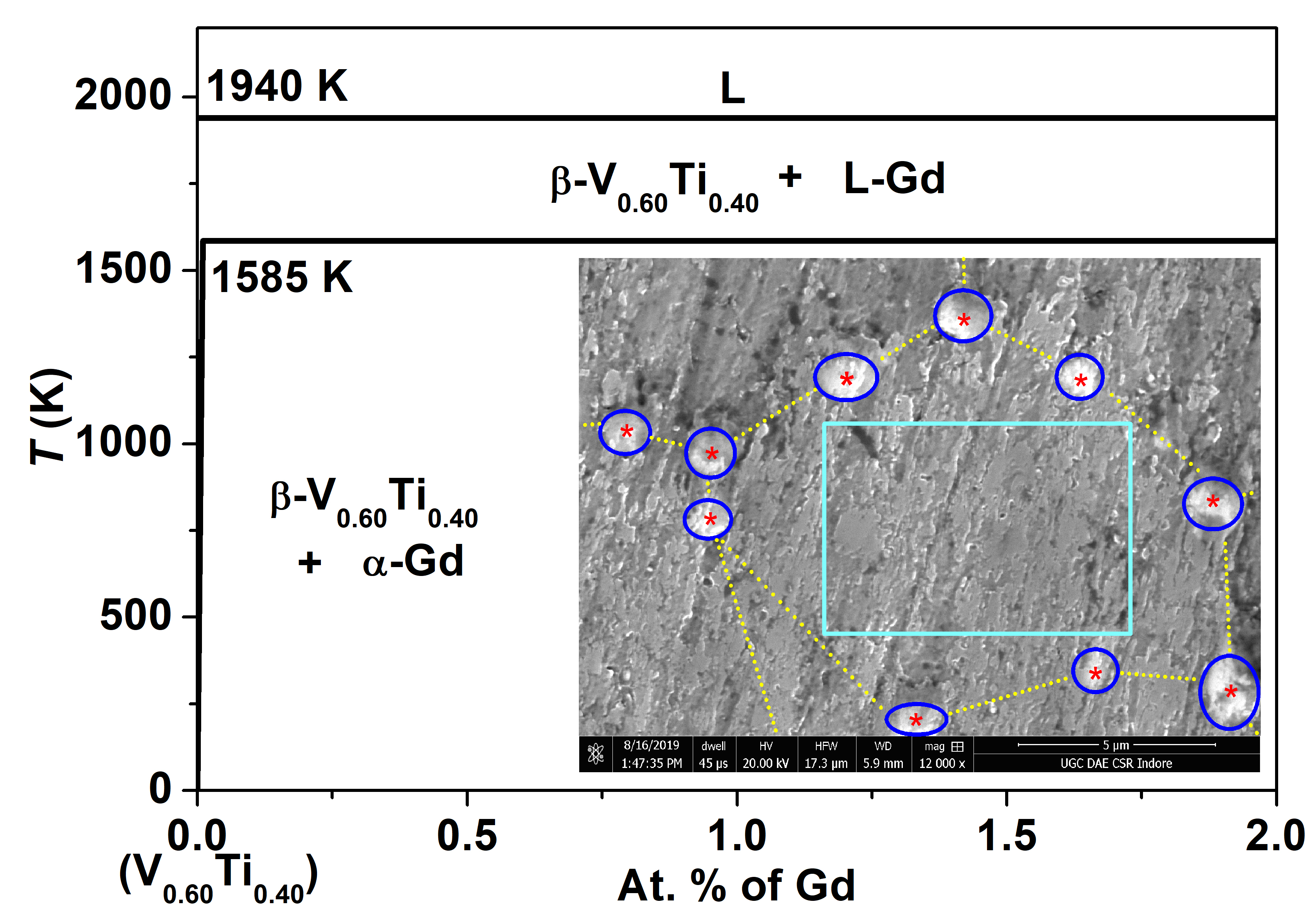}
 	\caption{Phase diagram and microstructure of the (V$_{0.60}$Ti$_{0.40}$)-Gd alloys. These alloys form homogeneous liquid above 1940~K and separate into $\beta$ and $\alpha$ phases. (Inset) Scanning electron microscopy images of the Gd2 alloy. Gd precipitates along the grain boundaries (marked by yellow dotted lines) with a dimension of about 1.2~$\mu$m (marked '$\star$'). No trace of Gd is observed inside the grains (cyan box).}
	\label{2}
 \end{figure}

Figure \ref{1} shows the XRD patterns of the (V$_{0.60}$Ti$_{0.40}$)-Gd alloys. Distinct peaks are seen for the reflections from the body centred cubic (bcc) V$_{0.60}$Ti$_{0.40}$ phase (Fig.\ref{1}(a)) and hexagonally close packed (hcp) Gd phase (Fig.\ref{1}(b)). The positions of the peaks corresponding to the V$_{0.60}$Ti$_{0.40}$ alloy do not shift with the amount Gd present in the sample. This indicates that the hcp-Gd is immiscible in the V$_{0.60}$Ti$_{0.40}$ alloy. The lattice parameter of the bcc phase is 3.13 \AA, which is in agreement with that reported  in literature for the V$_{0.60}$Ti$_{0.40}$ alloy \cite{mat15}. The lattice parameters of the hcp-Gd precipitates are $a$ = 3.636~\AA~and $c$ = 5.783~\AA. These values are similar to those reported in literature for bulk Gd \cite{gra09}. The presence of a few peaks with lower intensities and shifted positions corresponding to the reflections from hcp-Gd indicates that there is a distribution in the sizes of Gd clusters.     

The rare earth elements are immiscible in solid vanadium \cite{cha10, lov60, gschneidner64, kom60, col59, pen17, smi88,  bus77}. It was predicted theoretically that no binary Ti-rare earth and V-rare earth compounds will form \cite{bus77}.  However, there is a solubility of rare earths in the liquid vanadium when the solute concentration is low \cite{cha10, lov60, col59, smi88}. The range over which a homogeneous liquid forms increases with increasing temperature \cite{lov60,kom60,cha10}. Similar behaviour is also observed in Ti-rare earth alloys \cite{lov60}. The SEM images of (V$_{0.60}$Ti$_{0.40}$)-Gd alloys \cite{paul20} reveal that fine clusters of Gd precipitates along the GBs. The size of the precipitates is $<$ 1.2~$\mu$m (marked by '$\star$' for the Gd2 alloy in the inset to Fig. \ref{2}). Precipitation of fine clusters requires solubility at least in the liquid phase \cite{sis01}. The phase diagram of the (V$_{0.60}$Ti$_{0.40}$)-Gd alloys up to 2 at.\% Gd shown in Fig. \ref{2} is constructed by comparing the phase diagrams of V-Gd \cite{cha10, lov60} and Ti-rare earth \cite{lov60} with the metallography images of the present alloys. The V$_{0.60}$Ti$_{0.40}$ alloy melts at about 1940~K above which the (V$_{0.60}$Ti$_{0.40}$)-Gd alloys up to 2 at.\% of Gd form homogeneous liquid phase. The solubility of Gd in the solid $\beta$-V$_{0.60}$Ti$_{0.40}$ phase is negligible. On the other hand, Gd remains in liquid phase above 1585~K. Therefore, when cooled below 1940~K, the homogeneous (V$_{0.60}$Ti$_{0.40}$)-Gd solution phase separates into solid $\beta$-V$_{0.60}$Ti$_{0.40}$ phase and liquid Gd. The extrusion of the liquid Gd from the solid $\beta$-V$_{0.60}$Ti$_{0.40}$ phase results in the formation of fine droplets of liquid Gd. These droplets generate stain field which hinders the growth of the $\beta$-V$_{0.60}$Ti$_{0.40}$ grains. The grain size of the Gd2 sample is about 20-30~$\mu$m (inset to Fig.\ref{2}) which is few orders of magnitude smaller than Gd0 alloy  \cite{paul20}. There is no trace of Gd inside the grains (region marked by the cyan box). Reduction of the grain size in these alloys is found to be useful in improving the critical current density \cite{paul20}.  Immiscibility of Gd in V$_{0.60}$Ti$_{0.40}$ results in the retention of ferromagnetism of bulk Gd  in the present alloys \cite{paul20}. Therefore, we present in the next section, the role of microstructure on the magnetic properties of the (V$_{0.60}$Ti$_{0.40}$)-Gd alloys.  

\subsection{Magnetic properties of the (V$_{0.60}$Ti$_{0.40}$)-Gd alloys: Coexistence of superconductivity and ferromagnetism}

\begin{figure}[htb]
	\includegraphics[width=\linewidth, keepaspectratio]{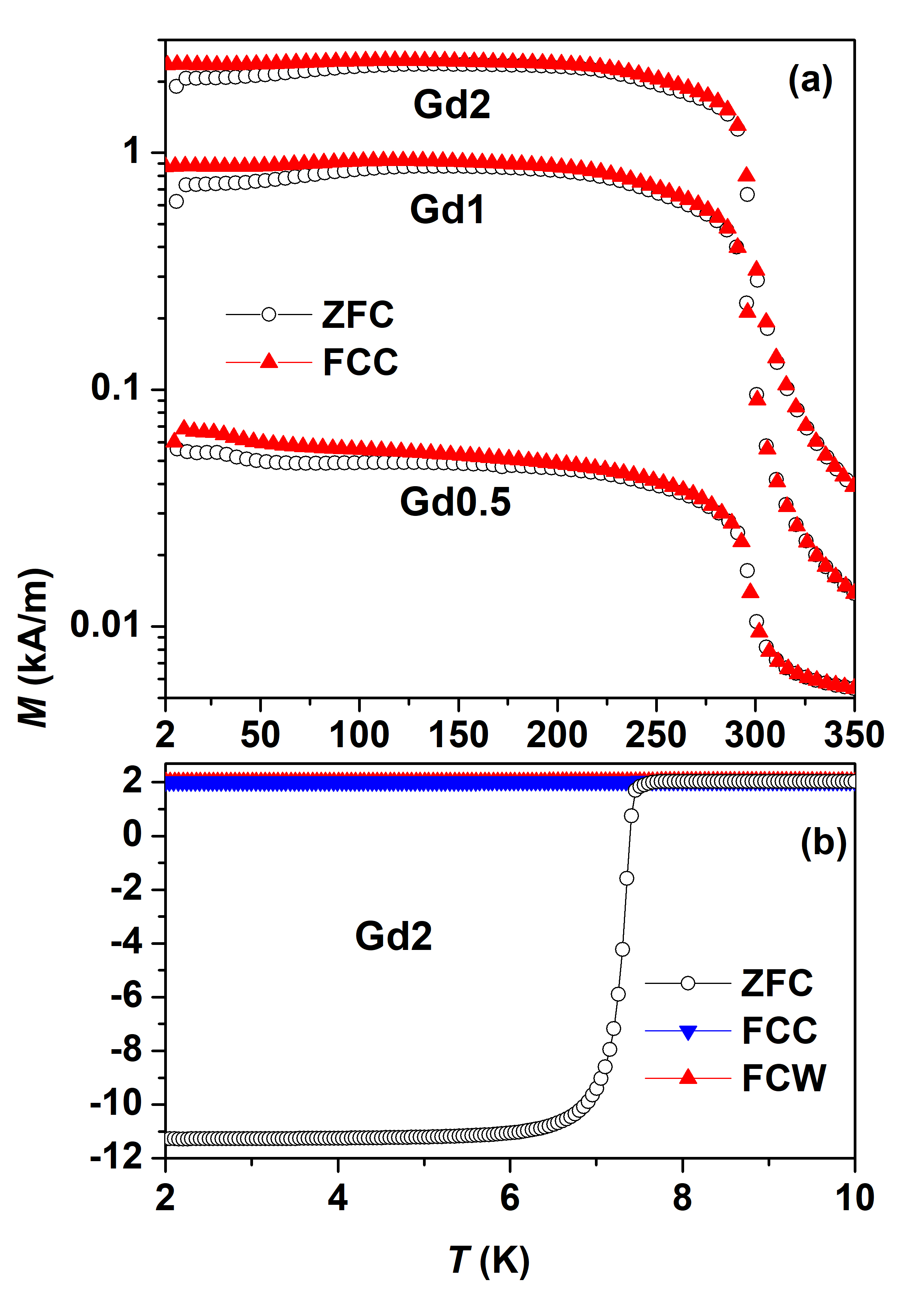}
	\caption{(a)  The $M$($T$) of the (V$_{0.60}$Ti$_{0.40}$)-Gd alloys measured in $H$ = 10~mT in the range 8-350~K. All the samples show the signature of a ferromagnetic transition at $T_{mc}$ = 295~K which is the $T_{mc}$ of elemental Gd. (b) The $M$($T$, H = 10~mT) below 10~K for the Gd2 sample depicts superconductivity below $T_{sc}$ = 7.75~K.}
	\label{3}
\end{figure}

\begin{figure}[htb]
	\includegraphics[width=\linewidth, keepaspectratio]{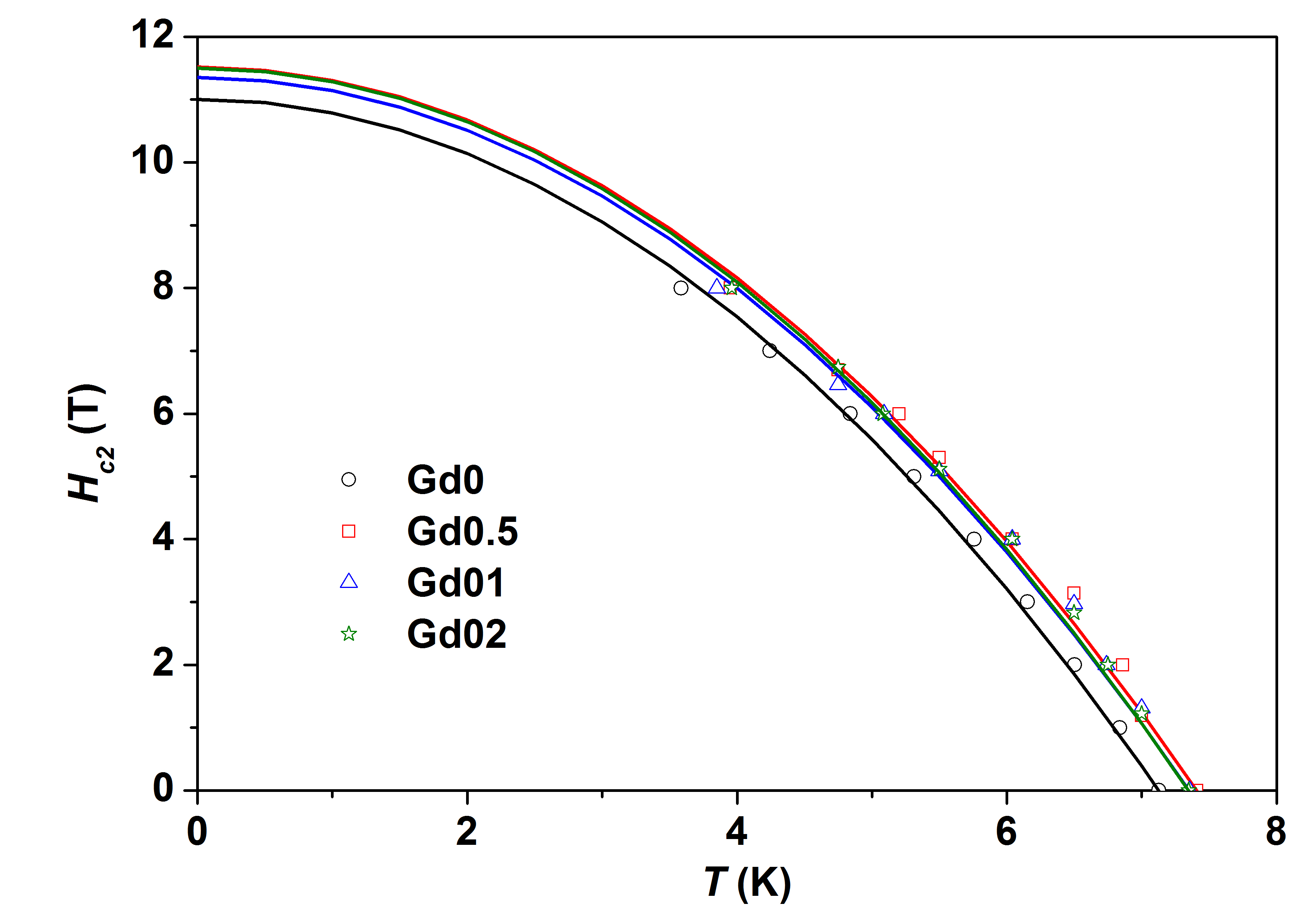}
	\caption{Temperature dependence of $H_{c2}$ of the (V$_{0.60}$Ti$_{0.40}$)-Gd alloys. The $H_{c2}$ of the Gd containing alloys at any given temperature is higher that that of the parent V$_{0.60}$Ti$_{0.40}$ alloy.}
	\label{10}
\end{figure}

 Figure \ref{3}(a) shows the temperature dependence of magnetization ($M$($T$)) of the (V$_{0.60}$Ti$_{0.40}$)-Gd alloys at $H$ = 10~mT in the range 8-350 K measured while warming after cooling down in zero field (ZFC) and while cooling after the ZFC measurement (FCC). All the samples show the signature of a paramagnetic to ferromagnetic transition at about $T_{mc}$ = 295~K which corresponds to that of elemental Gd \cite{dan98,paul20}. Figures \ref{3}(b) shows the $M$($T$) below 10~K for the Gd2 sample measured in the ZFC, FCC modes and while warming after the FCC measurement (FCW mode). A large decreasing $M$($T$) below 7.75~K is due to superconductivity. The large positive magnetization during FCC and FCW and in a large part of the ZFC curves indicates strong magnetic correlations within the superconducting state. 

\begin{figure}[htb]
	\includegraphics[width=\linewidth, keepaspectratio]{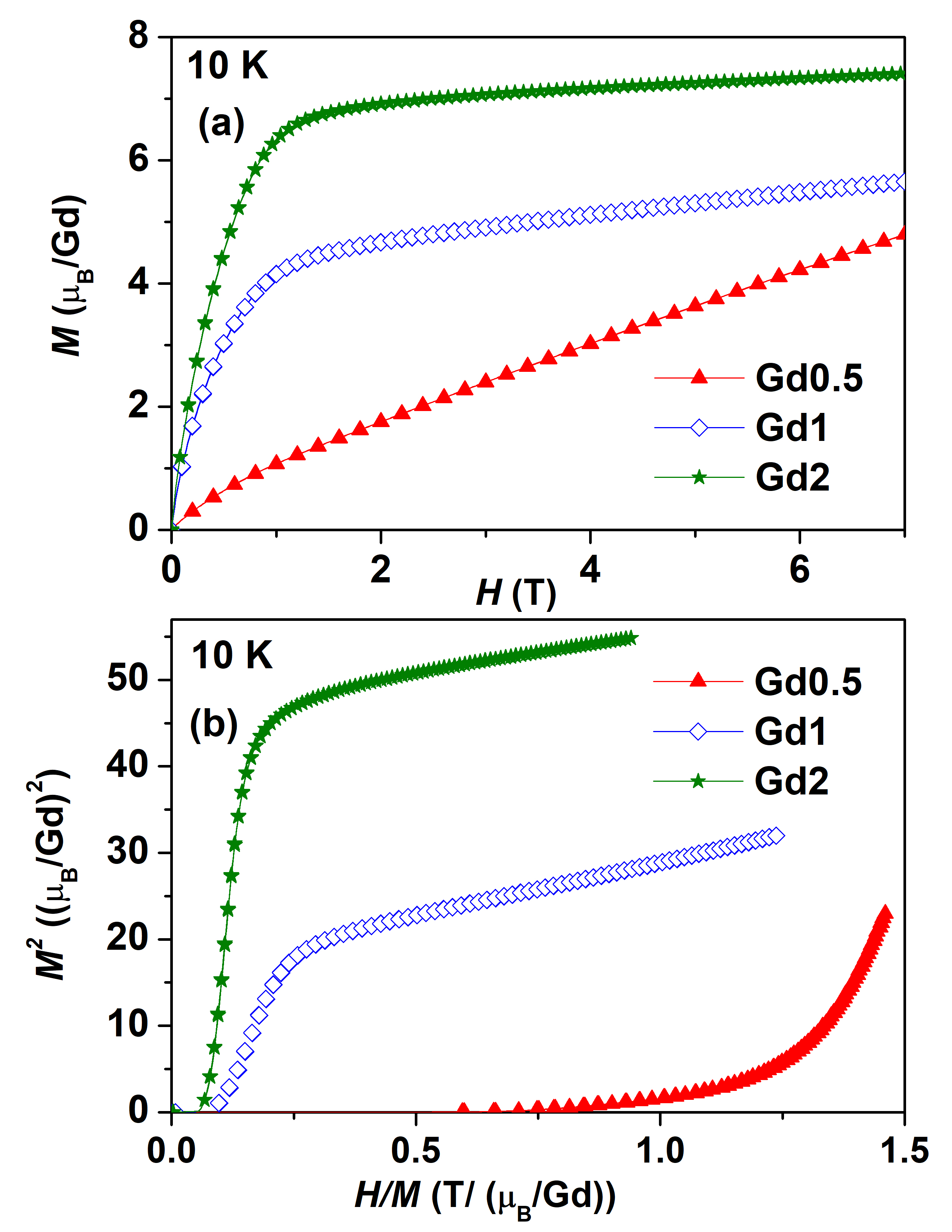}
	\caption{(a)  The $M$($H$) of (V$_{0.60}$Ti$_{0.40}$)-Gd alloys measured at 10~K. Magnetization saturates above 1~T in the samples containing 1 at.\% Gd or more. (b) The Arrott plot for the Gd containing alloys at 10~K. Long range ferromagnetism is established in samples containing 1 at.\% Gd or more.}
	\label{4}
\end{figure}

W have also measured the magnetic field dependence of magnetization ($M$($H$)) at various temperatures in the normal state as well as in the superconducting state \cite{paul20}.  The upper critical field ($H_{c2}$) is estimated as that magnetic field at which the magnetization deviates from the normal state value in high fields.  The $H_{c2}$ at different temperature was also estimated from the heat capacity measurement. Figure \ref{10} shows the temperature dependence of $H_{c2}$ for the (V$_{0.60}$Ti$_{0.40}$)-Gd alloys. The $H_{c2}$ of Gd containing alloys is found to be higher than that of the parent V$_{0.60}$Ti$_{0.40}$ alloy.  The $H_{c2}$($T$) follows the relation $H_{c2}(T) = H_{c2}(0)(1 - {(T/T_c)}^2)$ \cite{schmidt97}. The $H_{c2}(0)$ for the V$_{0.60}$Ti$_{0.40}$ alloy is 11 T and is about 11.5~T for the alloys containing Gd.

Figure \ref{4} shows the (a) normal state $M$($H$) and (b) Arrott plot at 10~K for the Gd containing alloys. Though there is a slight change of curvature in the low fields for the Gd0.5 alloy, the $M$($H$) increases almost linearly. For the Gd1 and Gd2 alloys at 10~K, the technical saturation is observed above 1~T. The presence of spontaneous magnetization ($M_s$) is confirmed using the Arrott plots for the Gd1 and Gd2 alloys at 10~K. The $M_s$ is about 4.05~$\mu_B$/Gd atom and 6.67~$\mu_B$/Gd atom for Gd1 and Gd2 alloys respectively. This indicates that there is long range and bulk ferromagnetism in the samples which contain 1 at.\% Gd  or more. For both these alloys, the $M_s$ is lower than that for elemental Gd (8~$\mu_B$/Gd atom). The $M_s$ increases with the increasing amount of Gd in the V$_{0.60}$Ti$_{0.40}$ alloy. This indicates that there is a polarization of conduction electrons by the Gd magnetic moments. Since, the magnetic moments of Gd are localized, the long range ferromagnetism can arise only from Ruderman-Kittel-Kasuya-Yosida (RKKY)  interaction \cite{blundell01}. Since, Gd clusters are along the GBs, bulk ferromagnetism can only arise from the conduction electron polarization along and around the GBs. Therefore, there is a competition between the incoherent scattering of electrons from the GBs and coherent scattering due to conduction electron polarization. To shed more light on this aspect, we present in the next section a detailed study on electrical and thermal properties of these alloys.  

\subsection{Role of microstructure on the electrical and thermal properties of (V$_{0.60}$Ti$_{0.40}$)-Gd alloys}

\begin{figure}[htb]
	\includegraphics[width=\linewidth,height=15 cm, keepaspectratio]{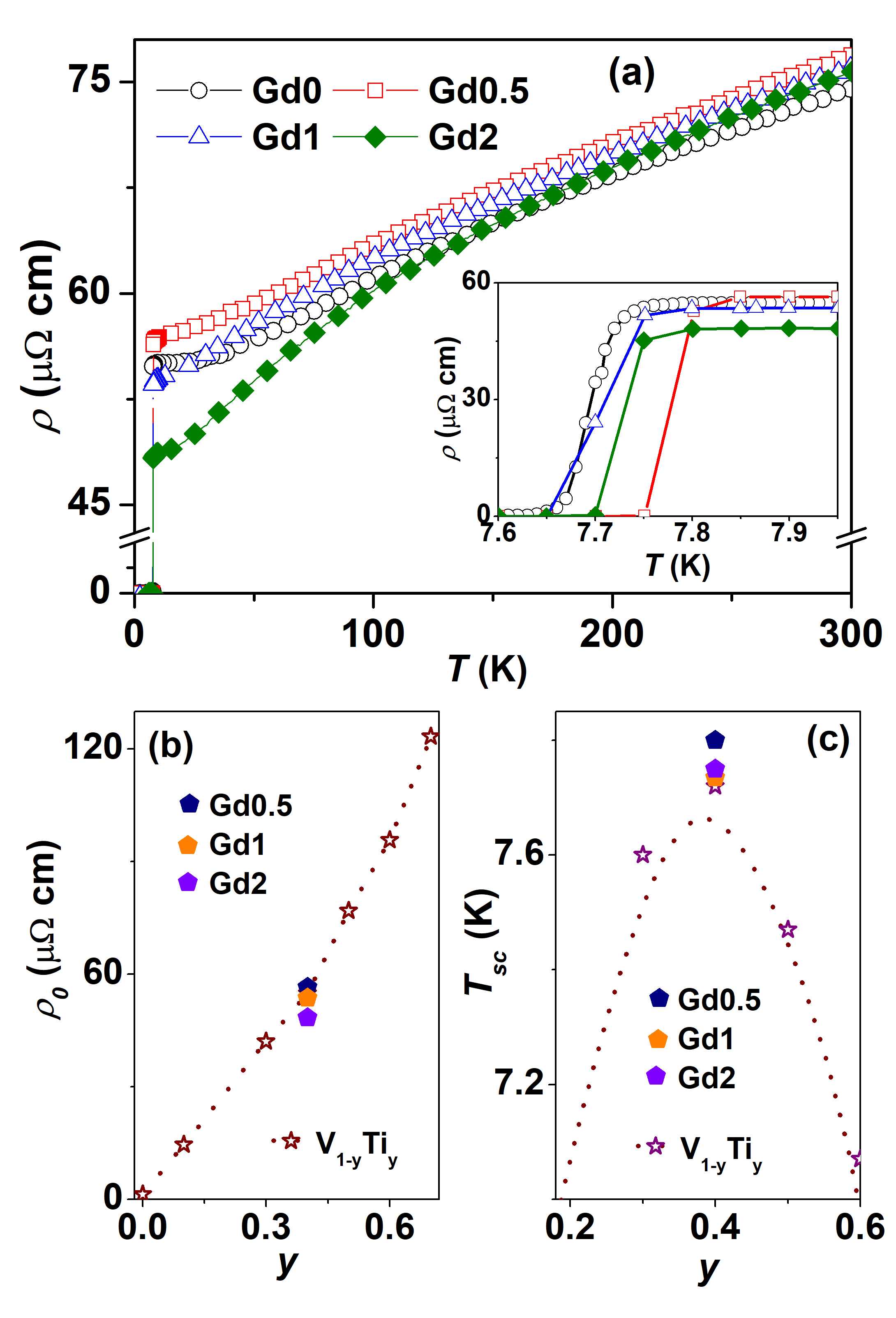}
	\caption{ (a) The $\rho(T)$ of (V$_{0.60}$Ti$_{0.40}$)-Gd alloys in the range 2-300 K. The $\rho(T)$ at $T < $100~K decreases with increasing amount of Gd above 0.5 at.\%. Extended view of $\rho(T)$ around $T_{sc}$ indicates that the $T_{sc}$ of Gd containing alloys is higher than that of the Gd0 alloy. (b) The $\rho_0$ of (V$_{0.60}$Ti$_{0.40}$)-Gd alloys is lower than that of V$_{1-y}$Ti$_{y}$ alloys. (c) The $T_{sc}$ of V$_{0.60}$Ti$_{0.40}$ alloy is the highest among the V$_{1-y}$Ti$_{y}$ alloys. The $T_{sc}$ of the Gd containing alloys is higher than all the V$_{1-y}$Ti$_{y}$ alloys.}
	\label{5}
\end{figure}

Figure \ref{5}(a) shows the $\rho$($T$) in the range 2- 300~K for the (V$_{0.60}$Ti$_{0.40}$)-Gd alloys in $H$ = 0. The residual resistivity ($\rho_0$) of the Gd0.5 alloy is slightly higher than that of the Gd0 alloy. The $\rho_0$ decreases with further increase in the amount of Gd in the V$_{0.60}$Ti$_{0.40}$ alloy. This indicates that the mean free path of electrons ($l_e$) increases in spite of the increase in disorder.  Figure \ref{5}(b) shows that the $\rho_0$ of the Gd containing alloys is lower than the $\rho_0$ of the V$_{y}$Ti$_{1-y}$ alloys having similar V and Ti compositions. Thus, the changes in the $\rho_0$ of (V$_{0.60}$Ti$_{0.40}$)-Gd alloys is not due to changes in the relative amounts of V and Ti. The increased GB density should have resulted in the increase of $\rho_0$ which is in contrast with the experimental observation. The variation of $\rho(T)$ around the $T_{sc}$ of the (V$_{0.60}$Ti$_{0.40}$)-Gd alloys shown in the inset to Fig. \ref{5}(a) indicates that the $T_{sc}$ of the Gd containing alloys is higher than that of Gd0 alloy. Fig. \ref{5}(c) shows that the $T_{sc}$ of the V$_{0.60}$Ti$_{0.40}$ alloy is the highest among all the V$_{y}$Ti$_{1-y}$ alloys. The (V$_{0.60}$Ti$_{0.40}$)-Gd alloys have $T_{sc}$ higher than the Gd0 alloy indicating that this variation is not due to the relative amounts of V and Ti present in the samples. 

The $\rho$($T$) of the Gd0 alloy above 200~K is lower than that of the Gd containing alloys. Therefore, the reduction of $\rho$($T$) at low temperatures in the alloys containing more than 0.5 at.\% Gd as compared to that of Gd0 alloy may not be due to ferromagnetism. On the other hand, the V$_{1-y}$Ti$_{y}$ alloys exhibit IFSF \cite{mat14ejpb, mat14}. Moreover, the fluctuation conductivity in the V$_{0.40}$Ti$_{0.60}$ alloy at $T < 3T_{sc}$ indicates a local variation of the FSF coupling constant ($\lambda_{sf}$) due to random disorder \cite{mat14}. As mentioned earlier, the addition of Gd suppresses the FSF in the uranium based systems such as UAl$_2$ and UAl$_3$ through the exchange interaction between the localized Gd 4$f$ moments and conduction electrons \cite{oli93, luc96, bur91}. In similar lines, the addition of Gd in the V$_{0.60}$Ti$_{0.40}$ alloy results in the suppression of IFSF. This is also supported by the fact that the $T_{sc}$ of the Gd containing alloys is more than that of the Gd0 alloy. This can be verified from the temperature dependence of thermal conductivity ($\kappa$($T$)), as the IFSF does not limit the heat conduction by the electrons and phonons \cite{paul19}. The changes in $\kappa$($T$) are mainly due to the scattering by the defects generated in the V$_{0.40}$Ti$_{0.60}$ alloy with the addition of Gd.

\begin{figure}[htb]
	\begin{center}
		
	\includegraphics[width=\linewidth, keepaspectratio]{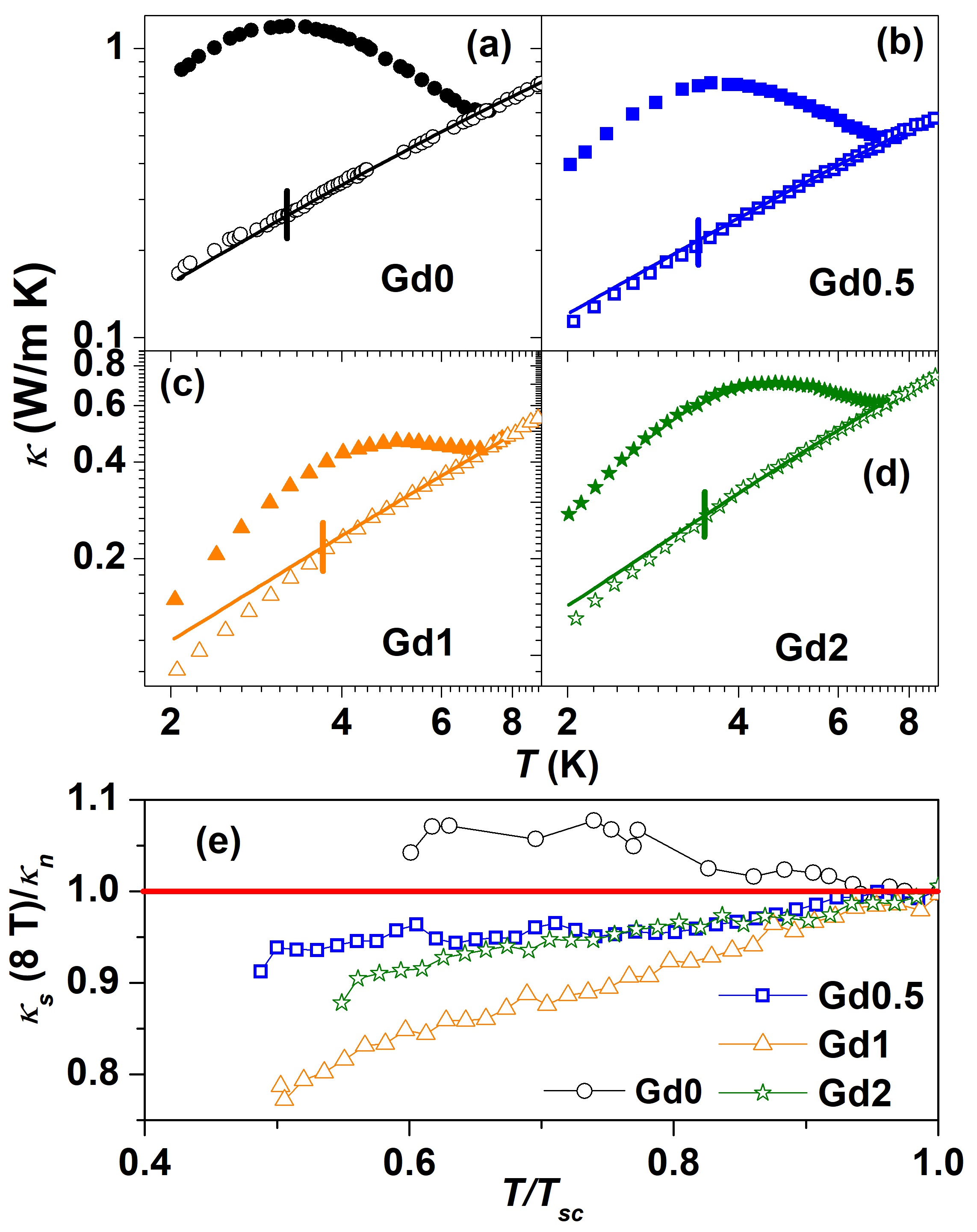}
	\caption{(a)-(d) Temperature dependence of thermal conductivity ($\kappa(T)$) of the (V$_{0.60}$Ti$_{0.40}$)-Gd alloys for zero field (filled symbols) and 8 T (open symbols) in the temperature range 2 - 9 K. (e) The ratio of  $\kappa_s$($H$ = 8 T) to $\kappa_n$  as a function of $T/T_{sc}$. In comparison with $\kappa_n$, the $\kappa_s$ increases below $T_{sc}$ for the Gd0 alloy, whereas it decreases for the Gd containing alloys.}
	\label{6}
	\end{center}
\end{figure}

The thermal conductivity of a superconducting material is given by \cite{paul19, tritt2004, chandra12}
\begin{equation}
\kappa_z = \kappa_{ez} + \kappa_{lz},
\end{equation}
where $\kappa_{ez}$ and $\kappa_{lz}$ are respectively the electronic and phononic thermal conductivities in the superconducting ($z$ = $s$) and normal ($z$ = $n$) states. The ${\kappa_{en}}^{-1} = {\kappa_{ei,n}}^{-1} + {\kappa_{el,n}}^{-1}$, where ${\kappa_{ei,n}}^{-1} = {A_n}T^{-1}$ and ${\kappa_{el,n}}^{-1} = {B_n}[T^2 +\mathcal{O}(T^4)]$ are the electronic thermal resistivities due to scattering of electrons from the defects and phonons respectively \cite{tritt2004}. The Cooper pairs being in zero entropy state do not contribute to $\kappa_s$. The electronic thermal conductivity due to the normal electrons in the superconducting state limited by impurity scattering ($\kappa_{ei,s}$) has been given by Bardeen-Rickayzen-Tewordt \cite{bardeen59,tewordt61,tewordt62} as

\begin{eqnarray}\nonumber
\frac{\kappa_{ei,s}}{ \kappa_{ei,n} } &=&R_{ei} \\
&=& \frac{1}{f(0)} [f(-y) + y ln(1 + exp(-y)) +\frac{y^2}{2(1+exp(y))}]
\end{eqnarray}

where the Fermi integral $f(-y)$ is given by $f(-y) = \int_{0}^{\infty} \frac{zdz}{1+exp(z+y)}$. The electronic thermal conductivity due to the normal electrons in the superconducting state limited by the scattering from phonons ($\kappa_{el,s}$)  is given by Tewordt as \cite{tewordt62, tew63},

\begin{eqnarray}\nonumber	
\frac{\kappa_{el,s}}{\kappa_{el,n}} &=& R_{el}\\
&=& \frac{\int^{\infty}_{1} dx x (x^2 - 1)^{0.5}sech^2(0.5yx)G(x)}{\int^{\infty}_{0} dz z^2 sech^2(0.5z)G_0(z)} ,
\end{eqnarray}

where G(x) and G$_0$(z) are integrals given by Tewordt \cite{tewordt62, tew63}. The $y=\frac{\Delta(T)}{k_B T}$, where $\Delta(T)$ is the superconducting energy gap and $k_B$ is the Boltzmann constant. The temperature dependence of the energy gap $\Delta(T)$ can be determined for an isotropic superconductor as $\Delta(T)$  = $\Delta_0$tanh$\{$1.82[1.018(T$_C$/T - 1)]$^{0.51}\}$ where $\Delta_0$ is the superconducting energy gap in the limit of absolute zero \cite{carrington03,sundar15}. Since, Gd is immiscible in the V$_{0.40}$Ti$_{0.60}$ alloy, we have used the $\Delta_0$  of the V$_{0.40}$Ti$_{0.60}$ alloy \cite{paul19} for all the Gd containing V$_{0.40}$Ti$_{0.60}$ alloys. 

\begin{figure}[htb]
	\begin{center}
	\includegraphics[width=\linewidth, keepaspectratio]{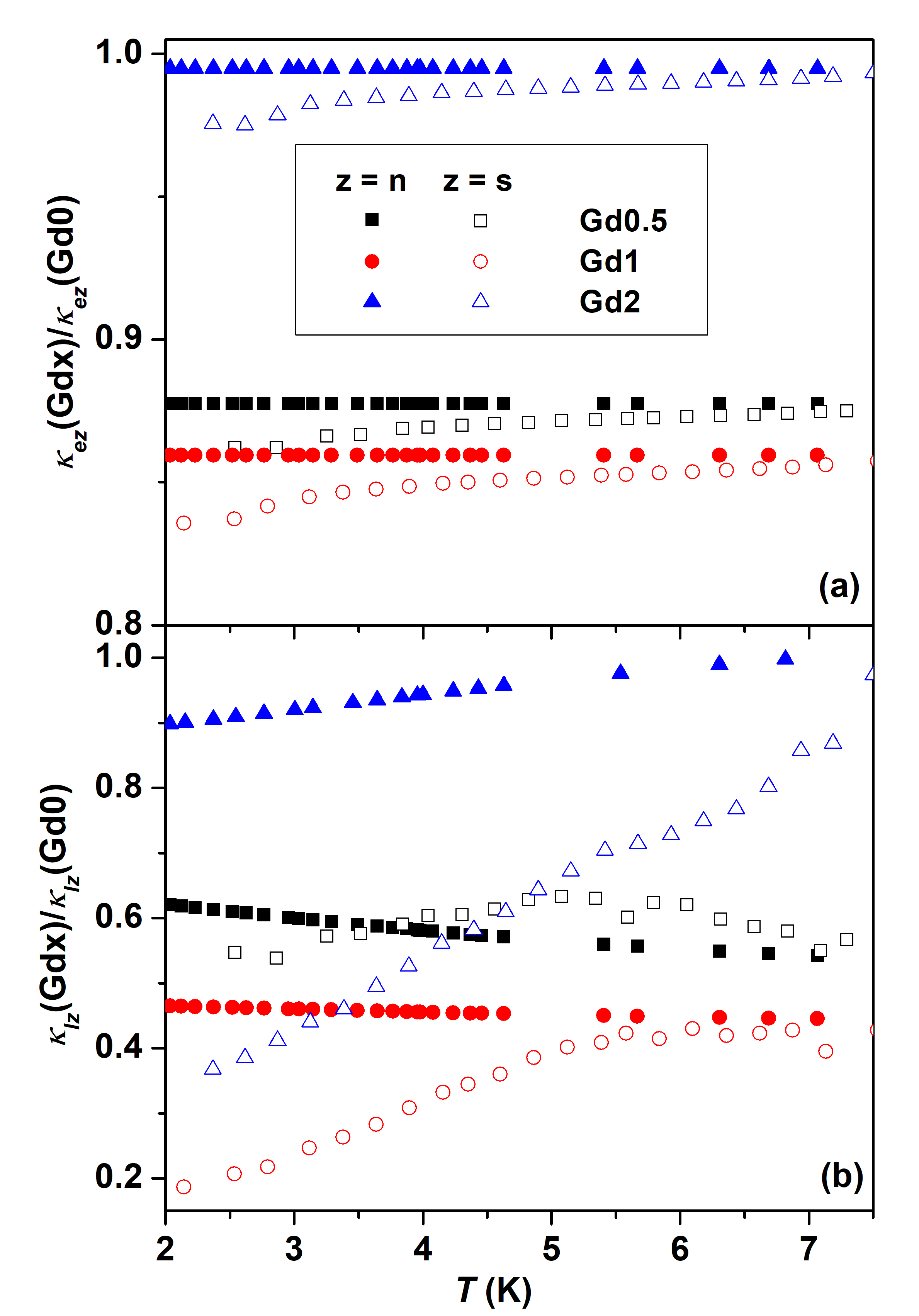}
	\caption{Ratios of (a) $\kappa_e(x)/\kappa_{e}(0)$, (b) $\kappa_l(x)/\kappa_{l}(0)$ as a function of temperature. Both $\kappa_{e}$ and $\kappa_{l}$ reduce for the alloys containing Gd. The $\kappa_{e}$ reduces at most 15\%, whereas the $\kappa_{l}$ reduces up to 80\% for the Gd1 alloy in comparison with that of the parent Gd0 alloy.}
	\label{7}
	\end{center}
\end{figure} 

The phonon thermal conductivity of superconducting materials is given as \cite{chandra12,kes74,tewordt89},
\begin{equation}
\kappa_{lz}(T) = M{T^3}\int_{0}^{\infty}du\frac{{u^4}{e^u}{({e^u}-1)^{-2}}}{N_z + L_zuT + C_zug(u)T + P_z{u^4}{T^4}},
\end{equation}
where $M = \frac{{k_B}^4}{2{\pi}^2{\hbar}^3v_s}$, $\hbar$ is reduced Plank's constant and $v_s$ is the sound velocity in the material. $N_z$, $L_z$, $C_z$ and $P_z$ are the phonon scattering coefficients from boundary, dislocations, electrons and point defects respectively. The function $g(u)$ gives the change in the electron-phonon scattering due to the formation of Cooper pairs and $u$ ($={\hbar\omega}/{k_B}T$) is the reduced phonon energy \cite{bardeen59}. The exact form of $g(u)$ below $T_{sc}$ is available in ref. \cite{bardeen59}. The $g(u)$ is unity for the normal state. 

\begin{figure}[htb]
	\begin{center}	
	\includegraphics[width=\linewidth, keepaspectratio]{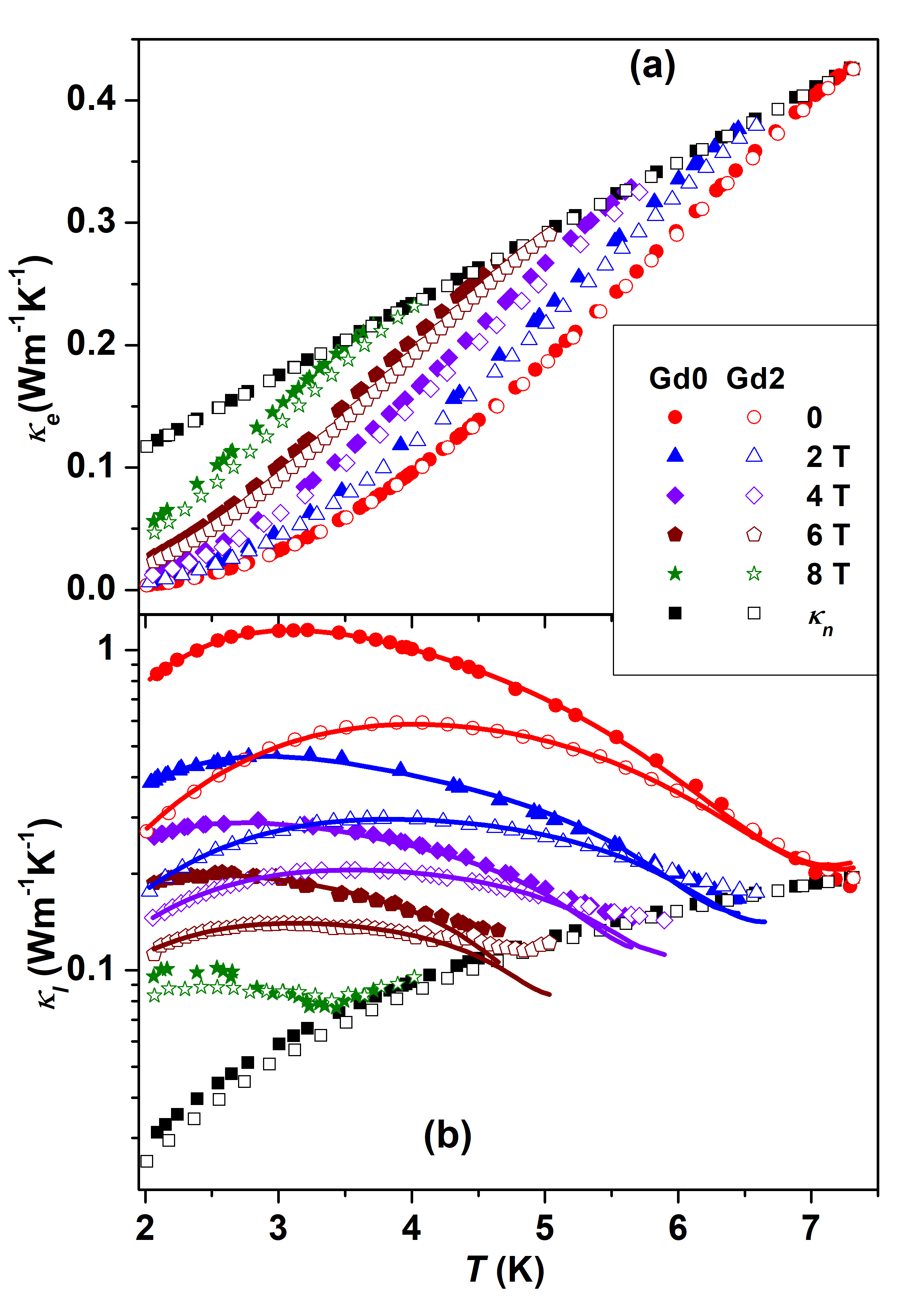}
	\caption{Comparison of (a) $\kappa_{es}(T)$ and (b) $\kappa_{ls}(T)$ of the Gd0 and Gd2 alloys in different magnetic fields. The $\kappa_{es}(T)$ of both the alloys in each field are almost the same but the  $\kappa_{ls}(T)$ of the Gd2 alloy is significantly lower than that of Gd0 alloy. For 0-8~T, the symbols are experimental data points and the lines are fit to the  $\kappa_{ls}(T)$. The normal state data $\square$ and $\blacksquare$ are obtained from the fitting of the experimental normal state data in the range 4.5-50~K}
	\label{8}
\end{center} 
\end{figure}

The $\kappa(T)$s of the (V$_{0.60}$Ti$_{0.40}$)-Gd alloys in the temperature range 2 - 9 K for $H$ = 0 (filled symbols) and $H$ = 8 T (empty symbols) are shown in the Fig.\ref{6}(a) to Fig.\ref{6}(d). The $\kappa$($T$) of the Gd containing alloys is observed to be similar to that of the V$_{1-y}$Ti$_y$ alloys \cite{paul19}. We have earlier shown that the heat is carried mainly by the electrons in the normal state, while the phonons dominate the heat conduction in the superconducting state of the V$_{1-y}$Ti$_y$ alloys \cite{paul19}. The grain size of the (V$_{0.60}$Ti$_{0.40}$)-Gd alloys reduces with increasing $x$ and becomes about 20 $\mu$m for the Gd2 alloy \cite{paul20}. However, the average phonon mean free path ($l_{ph}$) in these alloys is less than 100 nm \cite{paul19}. The presence of very few dislocations in these alloys are inferred from the absence of etch pits in the metallography images. Therefore, similar to the V$_{1-y}$Ti$_{y}$ alloys, the GBs and dislocations do not significantly affect the thermal conductivity. Hence, we have only considered the scattering of electrons from defects and phonons and the scattering of phonons from electrons and point defects to fit the $\kappa_n$($T$). The straight lines in the Fig. \ref{6}(a) to Fig. \ref{6}(d) show the fit obtained using eq.1 and eq.4 (in the temperature range 4.5 - 50~K) to the normal state thermal conductivity ($\kappa_n$). The $\kappa_{n}$ in the temperature range 4.5 - 8~K is obtained by applying 8~T magnetic field. The $\kappa_n$ below 4.5~K is generated using the parameters obtained from the fitting. The parameters are listed in table 2. Similar to the V$_{1-y}$Ti$_y$ alloys, the electrons carry majority of the heat in the normal state and the $\kappa_e$ is limited by the scattering of electrons from the static defects. Although the disorder increases with increasing Gd content in the V$_{0.60}$Ti$_{0.40}$ alloys \cite{paul20}, the coefficient $A_n$ (which depicts the strength of electron-disorder scattering) decreases slightly for the Gd2 alloy. This indicates that the electron mean free path ($l_{e}$) increases in spite of the increased disorder. The point defects scatter the high frequency phonons effectively, leaving the low frequency phonons to be scattered by the electrons.  The $C_n/M$ and $P_n/M$ increases initially and then decreases with increasing amount of Gd. Like in the V$_{1-y}$Ti$_y$ alloys \cite{paul19}, the rise of $\kappa_{s}(T)$ below $T_{sc}$ for $H = 0$ indicates that phonons carry majority of the heat in the superconducting state \cite{bardeen59,esquinazi83, richardson92}. The $\kappa(T)$ for $H =$ 8~T starts to deviate from $\kappa_{n}$ around 4~K for all the alloys (marked by '$\vert$' in the Fig. \ref{6}(a) to  Fig. \ref{6}(d)). The difference between $\kappa_s$ (8~T) and $\kappa_{n}$ being very small, we show in Fig. \ref{6}(e), the ratios of $\kappa_s$ (8~T) to $\kappa_{n}$ as a function of $T/T_{sc}$(8~T). As $T$ decreases below $T_{sc} (8~T)$, this ratio starts to rise for the Gd0 alloy but decreases for the Gd containing alloys. Thus there is a crossover from phonon dominant heat conduction (for $H <$ 8~T) to electron dominant heat conduction (for $H \geq$ 8~T) for the alloys containing Gd.

Figure \ref{7} shows the ratios (a) $\kappa_{ez}(Gdx)/\kappa_{ez}(Gd0)$, and (b) $\kappa_{lz}(Gdx)/\kappa_{lz}(Gd0)$ for Gd0.5, Gd1 and Gd2 alloys. The $\kappa_{e}$ and $\kappa_{l}$  in the both normal and superconducting ($H = 0$) states are obtained using eqs.(1-4). We see that the ratios $\kappa_{en}(Gd0.5)/\kappa_{en}(Gd0)$ and  $\kappa_{en}(Gd1)/\kappa_{en}(Gd0)$ decrease due to increased disorder. However, this ratio ($ \kappa_{en}(Gd2)/\kappa_{en}(Gd0)$) for the Gd2 alloy reaches almost unity. The $\kappa_{es}(Gdx)/\kappa_{es}(Gd0)$ follows the $\kappa_{en}(Gdx)/\kappa_{en}(Gd0)$. The ratio $\kappa_{ln}(Gdx)/\kappa_{ln}(Gd0)$ decreases up to 1 at.\% Gd and approaches unity for the Gd2 alloy, which is also similar to that observed in the electronic thermal conductivity. These results suggest that the electron-phonon interaction governs the thermal conductivity of these alloys. The $\kappa_{ls}(T)$ is estimated by subtracting $\kappa_{es}(T)$ from $\kappa_s(T)$. It is observed that the change in $\kappa_{ls}(Gdx)/\kappa_{ls}(Gd0)$ between $T_{sc}$ and $T =$ 2 K (Fig.7(b)) increases with increasing amount of Gd. This is due to the increase in the static defects which are effective scatterers of phonons in the superconducting state. We have observed that near 2~K, the $\kappa_{ln}$ ($\kappa_{ls}$) reduces by 50\% (80\%) for the Gd1 alloy  in comparison with the V$_{0.60}$Ti$_{0.40}$ alloy but $\kappa_{e}$ reduces only about 15\%. In the V$_{0.60}$Ti$_{0.40}$ alloy, apart from the scattering of phonons by the electrons and point defects, the phonons in the superconducting state are also scattered by the dislocations \cite{paul19}. Therefore, the larger reduction in $\kappa_{ls}$ in comparison with the $\kappa_{ln}$ of the Gd containing alloys is due to the additional dislocations generated by the addition of Gd.  

\begin{table}[h]
	
	\begin{center}
			
			\caption{Parameters obtained from the fitting of $\kappa_n(T)$ and $\kappa_{ls}(T)$}
			\label{tab:table1}
			\begin{tabular}{c|c|c|c} 
				\hline
				\multicolumn{4}{c}{\textbf{Normal state}}\\
				\hline
				\hline
				$x$ & $A_n$ & $C_n/M$ & $P_n/M$\\
				
				{} & (m K$^{2}$ W$^{-1}$)  & (m K$^3$ W$^{-1}$) & (m W$^{-1}$)\\
				\hline
				Gd0 & 15.9584 & 907.1563 & 0.1457\\
				
				Gd0.5 & 18.8599 & 1126.9 & 0.4028\\
				
				Gd1 & 19.9622 & 2607.4 & 0.1771\\
				
				Gd2& 17.3044 & 1106 & 0.0628\\
				\hline
				\hline
				\multicolumn{4}{c}{\textbf{Superconducting state}}\\
				\hline
				\hline
				$H$ & $C_s/M$ & $L_s/M$ & $P_s/M$\\
				(T) & (m K$^3$ W$^{-1}$)  & (m K$^3$ W$^{-1}$) & (m W$^{-1}$)\\
				\hline
				\multicolumn{4}{c}{\textbf{Gd0}}\\
				\hline
				
				\textbf{0} & 1607.5 & 31.5824 & 0.0041\\
				
				\textbf{2} & 1166.5 & 45.6484 & 0.074\\
				
				\textbf{4} & 900.4585 & 56.0613 & 0.1764\\
				
				\textbf{6} & 642.20 & 59.6212 & 0.3635\\
				\hline
				
				\multicolumn{4}{c}{\textbf{Gd2}}\\
				\hline
				
				\textbf{0} & 1892.6 & 95.8641 & 0.0084\\
				
				\textbf{2} & 1160.5 & 116.9342 & 0.0912\\
				
				\textbf{4} & 967.6129 & 128.525 & 0.1862\\
				
				\textbf{6} & 783.9803 & 129.3814 & 0.3965\\
				\hline
				
			\end{tabular}
			
			
			
	\end{center}
\end{table}

Figure \ref{8} shows the temperature dependence of (a) $\kappa_{e}$ and (b) $\kappa_{l}$ for the Gd0 and Gd2 alloys in different magnetic fields. We have analysed the $\kappa_s(T, H \neq 0)$ using the effective gap model (EGM) \cite{dubeck63,dubeck64}. According to the EGM, the effect of magnetic field on the $\kappa_s(T)$ can be incorporated as the field dependence of the effective superconducting energy gap at different temperatures ($\Delta(T, H)$), where $\Delta(T,H)$ = $\Delta$($T$) $\times$ (1 - ($B/H_{c2}$($T$))$^{1/2}$. Here $B \simeq H$ for $H$ $\gg$ $H_{c1}$. The $\kappa_{es}(T)$ and $\kappa_{ls}(T)$ in different applied magnetic fields are estimated using eqs.(1-4) and $H_{c2}(T)$ (shown in Fig.\ref{10}). The values of $\kappa_{es}$ for the Gd2 alloy is almost the same as that of Gd0 alloy for each $H$ as $\kappa_{en}(Gd2) \simeq \kappa_{en}(Gd0)$ at all $T$. However, $\kappa_{ls}$ of the Gd2 alloy is significantly lower in comparison with that of the Gd0 alloy. 

\begin{figure}[htb]
	\begin{center}	
	\includegraphics[width=\linewidth, keepaspectratio]{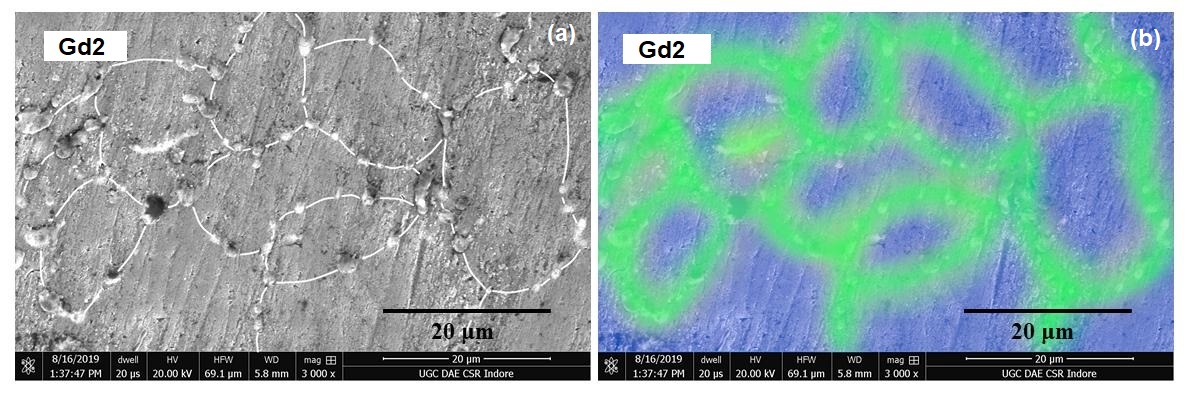}
	\caption{(a) Metallography image of the Gd2 alloy. (b) Reconstructed image of (a) depicting regions of superconductivity (blue) and regions of conduction electron polarization leading to ferromagnetism (green).}
	\label{9}
\end{center} 
\end{figure}

The solid lines in Fig. \ref{8}(b) are fits to the $\kappa_{ls}$ obtained using eq. (4). The fitting parameters are presented in table 2. The $C_s/M (H = 0)$ is higher than $C_n/M$, whereas $P_s/M$ is much lower than $P_n/M$ for both the alloys. The loss of $n_e$ in $H = 0$ below $T_{sc}$ increases the $l_{ph}$ of the low frequency phonons. This leads to the reduction in the scattering of these phonons from the point defects. Thus, the scattering of phonons from electrons governs the heat conduction below $T_{sc}$. As the temperature is further reduced, the $l_{ph}$ increases further and becomes comparable to the inter-dislocation distances. As a result, the dislocations now become the dominant scatterers of phonons. As $H$ is increased, the $n_e$ at any given temperature below $T_{sc}$ is higher than that at $H = 0$. Therefore, $l_{ph}(T, H \neq 0)$ is less than $l_{ph}(T, H = 0)$. Thus, the point defects and dislocations become more effective scatterers with increasing $H$. Hence, $P_s/M$ and $L_s/M$ increase with increasing $H$. On the other hand, the scattering of phonons from electrons becomes less effective with increasing $H$ which leads to lower $C_s/M$. Once the $H$ is sufficiently large, the $l_{ph}$ becomes quite smaller than the inter-dislocation distances. Therefore, the scattering of phonons from the electrons and point defects becomes more effective in limiting the heat flow. Additionally, the number of normal electrons being closer to the normal state value, the electrons become the major scatterers of phonons. The  $C_s/M$ and $P_s/M$ for each $H$ are almost the same for both the Gd0 and Gd2 alloys. The only difference is in the $L_s/M$, which is almost 2 - 3 times higher for the Gd2 alloy in comparison with the Gd0 alloy for each $H$. This indicates that the dislocations are one of the major disorders in the Gd containing V$_{0.60}$Ti$_{0.40}$ alloys. These dislocations pin the flux lines in the mixed state and contribute significantly to the enhancement of $J_c$ in these alloys \cite{paul20}. 

\subsection{Pictorial model: Evolution of physical properties in the (V$_{0.60}$Ti$_{0.40}$)-Gd alloys}

The understanding that evolved from the experimental results are summarized in Fig.\ref{9}. In this figure, (a) shows a large area SEM image of the Gd2 alloy. We see that the Gd precipitates are uniformly present along the GBs of the alloy. The conduction electron polarization along and around the GBs leading to a network of paths of  (RKKY) interaction is represented by the green regions in Fig. \ref{9}(b). The conduction electrons deep inside the grains (blue regions in Fig. \ref{9}(b)) only feel the average internal field generated by the magnetic ordering below $T_{mc}$. Thus, the IFSF are suppressed partially inside the grains resulting in the enhancement of the $T_{sc}$ in the Gd containing alloys. Increase in the disorder by the addition of Gd reduces the mean free path ($l_e$) of the conduction electrons, whereas the conduction electron polarization along the GBs due to Gd magnetic moments increases $l_e$. Therefore, effective $l_e$ decreases initially with the increase in the amount of Gd in the V$_{0.60}$Ti$_{0.40}$ alloy and then increases for larger amount of Gd. This results in the reduction of $\rho$($T$) below 100~K for the alloys containing more than 0.5 at.\% Gd and an enhancement of $\kappa$($T$) of the alloys containing more than 1 at.\% Gd.

\section{Conclusion}

The rare earth element Gd precipitates in the V$_{0.60}$Ti$_{0.40}$ alloy with a dimension $<$ 1.2~$\mu$m. This precipitation results in larger grain boundary and dislocation densities in comparison with those of the parent alloy. The precipitation also causes polarization of the conduction electrons along and around the grain boundaries due to the Gd magnetic moments. The conduction electron polarization results in the following features:\\
(a) A ferromagnetic transition at $T_{mc}$ = 295~K.\\
(b) Partial suppression of spin fluctuations inside the grains leading to slight enhancement of $T_{sc}$\\
(c) Increase of $l_e$ leading to reduction of $\rho$($T$) below 100~K for the alloys containing more than 0.5 at.\% Gd and an enhancement of $\kappa$($T$) of the alloys containing more than 1 at.\% Gd.

\section{Acknowledgement}

We thank Dr. R. Venkatesh, UGC-DAE Consortium for Scientific Research for SEM and EDAX measurements and Dr. Archna Sagdeo, Raja Ramanna Centre for Advanced Technology for XRD measurements.

\section{ Data Availability Statement:} The data that support the findings of this study are available from the corresponding author upon reasonable request.




\section*{References}
\begin{thebibliography}{1}

		\bibitem{mat14} Md. Matin, L. S. Sharath Chandra, Radhakishan Meena, M. K. Chattopadhyay, A. K. Sinha, M. N. Singh, S. B. Roy, Spin-fluctuations in Ti$_{0.6}$V$_{0.4}$ alloy and its influence on the superconductivity, Physica B 436 (2014) 20-25.
		
		\bibitem{mat13} Md. Matin, L. S. Sharath Chandra, M. K. Chattopadhyay, R. K. Meena, Rakesh Kaul, M. N. Singh, A. K. Sinha, S. B. Roy, Magnetic irreversibility and pinning force density in the Ti-V alloys, J. Appl. Phys. 113 (2013) 163903.
		
		\bibitem{tai07} M. Tai, K. Inoue, A. Kikuchi, T. Takeuchi, T. Kiyoshi, Y. Hishinuma, Superconducting properties of V-Ti alloys, IEEE Trans. Appl. Supercond. 17 (2007) 2542-2544.
		
		\bibitem{mat15} Md. Matin, L. S. Sharath Chandra, M. K. Chattopadhyay, R. K. Meena, Rakesh Kaul, M. N. Singh, A. K. Sinha, S. B. Roy, Critical current and flux pinning properties of the superconducting Ti-V alloys, Physica C 512 (2015) 32-41.
		
		\bibitem{sha19} L. S. Sharath Chandra, Sabyasachi Paul, Ashish Khandewlwal, Vinay Kaushik, Archna Sagdeo, R. Venkatesh, Kranti Kumar, A. Banerjee, M. K. Chattopadhyay, Structural and magnetic properties of the as-cast V$_{1-x}$Zr$_{x}$ alloy superconductors, J. Appl. Phys. 126 (2019) 183905.

		\bibitem{cha10} W. Chan, M. C. Gao, O. N. Dogan, and P. King, Thermodynamic assessment of V-rare earth systems, J. Phase Equilib. Diffus. 31 (2010) 425-432.

		\bibitem{lov60} B. Love, The metallurgy of yttrium and the rare earth metals, WADD Technical report no. 60-74, part I, Wright-Patterson Airforce base, Ohio (1960). Permanent Online link: https://catalog.hathitrust.org/Record/009206837
	
		\bibitem{paul20} Sabyasachi Paul, SK. Ramjan, R. Venkatesh, L. S. Sharath Chandra, M. K. Chattopadhyay, Grain refinement and enhancement of critical current density in the V$_{0.60}$Ti$_{0.40}$ alloy superconductor with Gd addition, Accepted for publication in IEEE Trans. Appl. Supercond. doi: \url{10.1109TASC.2021.3052718}
		
		\bibitem{mat58} B. T. Matthias, H. Suhl, E. Corenzwit, Spin exchange in superconductors, Phys. Rev. Lett. 1 (1958) 92-94.

		\bibitem{wol15} C. T. Wolowiec, B. D. White, M. B. Maple, Conventional magnetic superconductors, Physica C, 514 (2015) 113-129.
		
		\bibitem{sch60} K. Schwidtal, Supraleitung aufgedampfter Bleischichten mit Zusatz von Gadolinium, Zeitschrift f\"{u}r Physik 158 (1960) 563-571.
		
		\bibitem{woo65} M. A. Woolf, F. Rief, Effect of Magnetic impurities on the density of states of superconductors, Phys. Rev. 137 (1965) 557-564.
		
		\bibitem{mat58a} B. T. Matthias, H. Suhl, E. Corenzwit, Spin exchange in superconductors, Phys. Rev. Lett. 1 (1958) 449-450.
		
		\bibitem{map68} M. B. Maple, The superconducting transition temparature of La$_{1-x}$Gd$_{x}$Al$_{2}$, Phys. Lett. 26A (1968) 513-514.
		
		\bibitem{mat14ejpb} Md. Matin, L. S. Sharath Chandra, S. K. Pandey, M. K. Chattopadhyay, S. B. Roy, The influence of electron-phonon coupling and spin fluctuations on the superconductivity of the Ti-V alloys, Eur. Phys. J. B. 87 (2014) 131(10 pp).
		
		\bibitem{bar14} J. Barker, D. Paul, A. Hillier, Spin fluctuations and their effect
		on superconductivity in	Titanium-Vanadium alloys, ISIS Student Meeting (2014)
		(external link: \url{ https://warwick.ac.uk/fac/sci/physics/
			research/condensedmatt/supermag/whoswho/joel_barker/
		isis_student_meeting_2014.pdf})
		
		\bibitem{oli93} N. A. deOliveira, A. A. Gomes, A. Troper, Rare earths in UAl$_{2}$: quenching of spin fluctuations?, Hyperfine Interactions 80 (1993) 1067-1070.
		
		\bibitem{luc96} P. Lucaci, E. Burzao, I. Lupsa, Magnetic properties of U$_{1-x}$Gd$_{x}$Al$_{3}$ and U$_{1-x}$Dy$_{x}$Al$_{3}$ systems, J. Alloys Compd. 238 (1996) L4-L6.
		
		\bibitem{bur91} E. Burzao, P. Lucaci,  Magnetic properties of U$_{1-x}$Gd$_{x}$Al$_{2}$ and U$_{1-x}$Dy$_{x}$Al$_{2}$ compounds, Solid State Commun. 79 (1991) 1077-1079.
		
		\bibitem{sin00} A. K. Sinha, A. Sagdeo, P. Gupta, A. Kumar, M. N. Sing, R. K. Gupta, S. R. Kane, S. K. Deb, Commissioning of angle dispersive X-ray diffraction beamline on Indus-2, AIP Conf. Proc. 1349 (2011) 503-504.
		
		\bibitem{gra09} S. Grazulis, D. Chateigner, R. T. Downs, A. T. Yokochi, M. Quiros, L. Lutterotti, E. Manakova, J. Butkus, P. Moeck, and A. Le Bail, Crystallography open database - an open access collection of crystal structures, J. Appl. Cryst. 42 (2009) 726-729.

		\bibitem{gschneidner64} K. A. Gschneidner, Jr., Structural and physical properties of alloys and intermetallic compounds, in: L. Eyring (Ed.), Progress in	the science and technology of rare earths, 1, 1964, Pergamon, New York pp. 222-258.

		\bibitem{kom60} A. S. Komjathy, R. H. Read and W. Rostoker, Phase relationships in selected binary and ternary vanadium-base alloys systems, WADD Technical report no. 59-483, Wright-Patterson Airforce base, Ohio (1960). Permanent Online link: https://catalog.hathitrust.org/Record/009230761

		\bibitem{col59} J. F. Collins, V. P. Calkins, and J. A. McGurty, Applications of rare earths to ferrous and non-ferrous alloys, American Society for Metals-Atomic Energy Commission Symposium on the Rare Earths and Related Metals, Chicago, Illinois (1959) doi:10.2172/4215576.

		\bibitem{pen17} L. Peng, C. Jiang, X. Li, P. Zhou, Y. Li, and X. Lai, The formation of precipitates and its effect on grain structure in V-4Cr-4Ti alloys with yttrium addition, J. Alloys Compd. 694 (2017) 1165-1174.

		\bibitem{smi88} J. F. Smith, K. J. Lee, and D. M. Martin, Binary rare earth-vanadium systems, CALPHAD 12 (1988) 89-96.

		\bibitem{bus77} K. H. J. Buschow, Intermetallic compounds of rare-earth and 3d transition metals, Rep. Prog. Phys. 40 (1977) 1179-1256.

		\bibitem{sis01} R. D. Sisson Jr., Heat treatment, in {\it Encyclopedia of Materials: Science and Technology} (eds. K. H. J. Buschow, M. C. Flemings, E. J. Kramer, P. Veyssiere, R. W. Cahn, B. Ilschner, and S. Mahajan Elsevier BV, Berlin (2001))

		\bibitem{dan98} S. Yu. Dan’kov, A. M. Tishin, V. K. Pecharsky, K. A. Gschneidner, Jr., Magnetic phase transitions and the magnetothermal properties of gadolinium, Phys. Rev. B 57 (1998) 3478-3490.

		\bibitem{schmidt97} V. V. Schmidt, The Physics of Superconductors, Springer, Verlag Berlin Heidelberg 1997.
		
		\bibitem{blundell01} S. Blundell, Magnetism in condensed matter, Oxford University Press, Oxford 2001.

		\bibitem{paul19} Sabyasachi Paul, L. S. Sharath Chandra, M. K. Chattopadhyay, Renormalization of electron-phonon coupling in the Mott-Ioffe-Regel limit due to point defects in the V$_{1-x}$Ti$_{x}$ alloy superconductors, J. Phys.: Condens. Matter 31 (2019) 475801.

		\bibitem{tritt2004} T. M. Tritt, Thermal Conductivity: Theory, Properties and Applications, Kluwer Academic / Plenum Publishers, New York 2004.
	 	
	 	\bibitem{chandra12} L. S. Sharath Chandra, M. K. Chattopadhyay, S. B. Roy, V. C. Saini, G. R. Myneni, Magneto thermal conductivity of superconducting Nb with intermediate level of purity, Supercond. Sci. Technol. 25 (2012) 035010.
	 	
	 	\bibitem{bardeen59} J. Bardeen, G. Rickayzen, L. Tewordt, Theory of thermal conductivity of superconductors, Phys. Rev. 113 (1959) 982-994.
	 	
	 	\bibitem{tewordt61} L. Tewordt, Lifetimes of quasi-particles and phonons in a superconductors at zero temperature, Phys. Rev. 127 (1961) 371-382.
	 	
	 	\bibitem{tewordt62} L. Tewordt,  Lifetimes of a quasi-particle in a superconductor at finite temperatures and application to the problem of thermal conductivity, Phys. Rev. 128 (1962) 12-20.
	 	
	 	\bibitem{tew63} L. Tewordt,  Theory of the intrinsic thermal conductivity of superconductors, Phys. Rev. 129 (1963) 657-663.
		
		\bibitem{carrington03} A. Carrington, F. Manzano, Magnetic penetration depth of MgB$_{2}$ Physica C 385 (2003) 205-214.
		
		\bibitem{sundar15} S. Sundar, L. S. Sharath Chandra, M. K. Chattopadhyay, S. B. Roy, Evidence of multiband superconductivity in the $\beta$-phase Mo$_{1-x}$Re$_{x}$ alloys, J. Phys. Condens. Matter. 27 (2015) 045701.
		
		\bibitem{kes74} P. H. Kes, J. G. A. Rolfes, D. de Klerk, Thermal conductivity of Niobium in the purely superconducting and normal states, J. Low Temp. Phys. 17 (1974) 341-364.
		
		\bibitem{tewordt89} L. Tewordt, and T. W\"{o}lkhaushen, Theory of thermal conductivity of the lattice for high-$T_{c}$ superconductors, Solid State Commun. 70 (1989) 839-844.
		
		\bibitem{esquinazi83} P. Esquinazi, F. de la Cruz, Phonon thermal conductivity limited by electron scattering in amorphous Zr$_{70}$Cu$_{30}$, Phys. Rev. B 27 (1983) 3069-3072.
		
		\bibitem{richardson92} R.A. Richardson, S. D. Peacor, C. Uher, F. Nori, YBa$_2$Cu$_3$O$_{7-\delta}$ films: calculation of the thermal conductivity and phonon mean-free path, J.Appl. Phys. 72 (1992) 4788-4791.
		
		\bibitem{dubeck63} L. Dubeck, P. Lindenfeld, E. A. Lynton, H. Rohrer, Thermal conductivity of superconductors of the second kind, Phys. Rev. Lett. 10 (1963) 98-101.
		
		\bibitem{dubeck64} L. Dubeck, P. Lindenfeld, E. A. Lynton, H. Rohrer, Magnetic and thermal properties of second-kind superconductors. II. Thermal conductivity in the mixed state, Rev. Mod. Phys. 36 (1964) 110-112.
		

		




\end{thebibliography}


\end{document}